# Advancing Risk Gene Discovery Across the Allele Frequency Spectrum

**Authors:** Madison Caballero[1,2], Behrang Mahjani[1-7]

**Corresponding author:**

Behrang Mahjani, PhD, One Gustave L. Levy Place, Box 1230, New York, NY 10029, behrang.mahjani@mssm.edu

**Affiliations:**

[1]Seaver Autism Center for Research and Treatment, Icahn School of Medicine at Mount Sinai, New York, NY, USA

[2]Department of Psychiatry, Icahn School of Medicine at Mount Sinai, New York, NY, USA

[3]Department of Genetics and Genomic Sciences, Icahn School of Medicine at Mount Sinai, New York, NY, USA

[4]Department of Artificial Intelligence and Human Health, Icahn School of Medicine at Mount Sinai, New York, NY, USA

[5]Mindich Child Health and Development Institute, Icahn School of Medicine at Mount Sinai, New York, NY, USA

[6]Department of Molecular Medicine and Surgery, Karolinska Institutet, Stockholm

[7]Department of Medical Epidemiology and Biostatistics, Karolinska Institutet, Stockholm, Sweden

## Abstract

The discovery of genetic risk factors has transformed human genetics, yet the pace of new gene identification has slowed despite the exponential expansion of sequencing and biobank resources. Current approaches are optimized for the extremes of the allele frequency spectrum: rare, high-penetrance variants identified through burden testing, and common, low-effect variants mapped by genome-wide association studies. Between these extremes lies variants of intermediate frequency and effect size where statistical power is limited, pathogenicity is often misclassified, and gene discovery lags behind empirical evidence of heritable contribution. This "missing middle" represents a critical blind spot across disease areas, from neurodevelopmental and psychiatric disorders to cancer and aging. In this review, we organize strategies for risk gene identification by variant frequency class, highlighting methodological strengths and constraints at each scale. We draw on lessons across fields to illustrate how innovations in variant annotation, joint modeling, phenotype refinement, and network-based inference can extend discovery into the intermediate range. By framing the frequency spectrum as a unifying axis, we provide a conceptual map of current capabilities, their limitations, and emerging directions toward more comprehensive risk gene discovery.

**Introduction**

The past two decades have witnessed remarkable progress in unraveling the genetic architecture of human disorders. Thousands of variants contributing to disease risk have been identified, from high-penetrance mutations underlying monogenic disorders to common variants with small individual effects revealed by genome-wide association studies (GWAS). Despite this progress and exponential growth in genomic datasets, the pace of new risk gene discovery for disorders has plateaued. Between 2011 and 2024, novel gene–disease associations in DisGeNET, a database that integrates evidence from curated sources, literature, and experiments, declined by more than 90% (**Fig 1A**) [1]. OMIM, which catalogs highly penetrant Mendelian disorders and their causal variants, reported a similar deceleration, although its scope is restricted to such variants [2]. This slowing does not reflect exhaustion of the genetic landscape as numerous genetic factors contributing to risk remain undiscovered. Rather, it signals that current methods have reached saturation, capturing most variants that are detectable under prevailing statistical frameworks, leaving a potential blind spot for variants that fall outside these detection limits.

To evaluate empirically whether such a blind spot exists, we analyzed ClinVar-classified missense variants within the 1000 Genomes Project cohort (1kGP) [3]. If detection were equally powered across the allele spectrum, variants without disease association would be disproportionately represented among common alleles. As expected, pathogenic annotations were strongly depleted among common alleles and modestly enriched at the rarest frequencies (**Fig 1B**). Unexpectedly, benign classifications, which lack any established disease relevance, peaked at intermediate frequencies (0.1–5%). Such an excess is inconsistent with neutral models. Most strikingly, benign calls were relatively more abundant in the 1–5% range than in the adjacent 5–10% range. Rather than representing a true enrichment of benign alleles, these findings likely reflect reduced statistical power to identify pathogenic effects outside the frequency extremes where current risk gene discovery methods are most sensitive.

Further evidence of the limits of conventional discovery methods is found in the many disorders with incomplete genetic explanations. Twin and family studies consistently demonstrate that complex disorders are profoundly genetic; autism spectrum disorder in particular approaches 80% heritability and many common disorders show substantial polygenic contributions [4,5]. Yet our ability to identify causal variants and risk genes falls dramatically short of these estimates. Autism is notable as, despite sequencing over 50,000 affected families, only 15-20% of autism cases yield an identifiable genetic cause. Missing genetic factors is also broadly apparent across somatic disorders like aging and cancer. For example, 10-20% of lung adenocarcinomas lack identifiable driver mutations which bars these patients from targeted therapies [6,7]. These missing genetic factors may reside in (1) classes of genetic variation that are technically difficult to detect or catalog (e.g., structural variants), and/or (2) in variants of lower penetrance, whether due to intermediate effect sizes or dependence on environmental exposures, developmental timing, or genetic background.

To address this gap, this review surveys emerging strategies for detecting both technically challenging variant types and variants with incomplete penetrance, particularly those with intermediate effect sizes and population frequencies. We focus on coding variation not because interpretation is straightforward—functional consequences, particularly of missense variants, remain challenging to assess—but because coding variants offer unambiguous gene-level mapping. A pathogenic coding variant in *BRCA2*, for example, directly implicates *BRCA2* itself. By contrast, regulatory variants may influence multiple genes across large genomic intervals. Thus, focusing on exonic variation allows us to isolate methodological constraints in risk gene identification, independent of locus mapping uncertainty.

Within this scope, we organize discovery strategies by allele frequency, highlighting how frequency and effect size interact to determine statistical power and biological interpretability. GWAS reliably identifies common variants with small effects, while burden tests successfully capture rare variants with large effects through aggregation. Between these well-powered extremes lies a broad and systematically underexplored frequency range: variants with a population-derived minor allele frequency (MAF) between 0.1% and 5%. These are typically underpowered for both GWAS and burden-based analyses, forming the aforementioned "missing middle" in genetic discovery.

By examining the spectrum of coding variation and integrating insights from cancer, psychiatric genetics, aging research, and select non-human studies, we highlight methodological innovations with broad cross-disciplinary relevance. Our goal is to delineate which genetic contributors are currently detectable, where systematic blind spots persist, and how the combination of new methods can extend the boundaries of risk gene discovery.

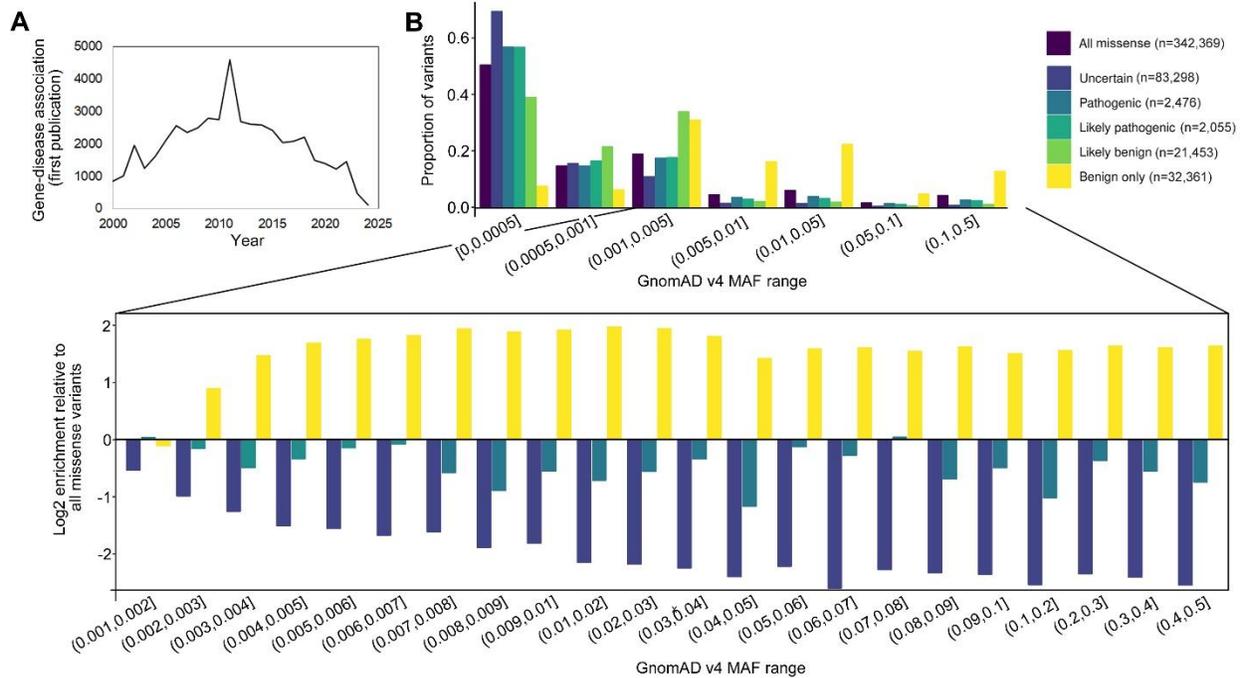

**Figure 1. Progress in gene–disease association discovery and representation of functional variation.** **(A)** Annual counts of novel gene–disease associations, based on first publication date, as catalogued in DisGeNET. **(B)** Top: Distribution of ClinVar variants (July 2025 release) present in the 1kGP cohort (n=3,202), stratified by GnomAD v4-derived minor allele frequency (MAF). Missense variants were defined in canonical transcripts and annotated with SIFT and PolyPhen-2, representing conservation-based and structure-informed predictors of missense variant impact, respectively. Variants were considered functional if they met thresholds of PolyPhen-2 ≥ 0.5 or SIFT ≤ 0.05. The minimum observed MAF was 0.000312 as all 1kGP variants are represented in GnomAD. Functional missense variants were then cross-referenced with ClinVar, and variants classified as benign were required to lack any conflicting non-benign annotations. Bottom: Expanded view of narrow MAF ranges for uncertain, pathogenic, and benign-only variants, with values expressed as the proportion of each ClinVar subset relative to the proportion of all functional missense variants within the range. Note the unexpected enrichment of benign variants at intermediate frequencies (0.1-5%), suggesting reduced power to detect pathogenic effects in this range.

# 1. Rare variation and large effect sizes

Rare variation, defined here as alleles with a MAF<0.1% and including *de novo* mutations, accounts for roughly half of genetic diversity within the human species but comprises <2% of variation within an individual [3]. This duality makes rare variants both plausible contributors to interindividual differences and a manageable subset for study. When they exert large phenotypic effects, rare variants are especially informative for risk gene mapping. In this section, we review the major classes of high-impact rare variation and their successes and limitations.

## 1.1 Copy number variations

Copy number variations (CNVs), encompassing deletions, duplications, and other dosage-altering events, are among the largest effect size variants and are implicated extensively in human disorders [8,9]. Microarray platforms reliably detect CNVs larger than 50kb and remain central to diagnostics and research, as pathogenic variants are typically large and rare owing to strong fitness effects [10,11]. Smaller CNVs will be discussed in the later section on common variation.

> **Highly penetrant CNVs impact multiple genes.** Large pathogenic CNVs often span multiple genes, complicating efforts to link genotype with phenotype. One strategy to refine interpretation is to study individuals with atypical breakpoints, when such variation exists, and compare clinical or molecular features according to the disrupted gene set. This approach has proven useful for identifying pathogenic drivers, and continues to be valuable in rare disease research, where phenotypes can often be aligned with specific disrupted loci [12].
>
> Recurrent rare CNVs are well-established contributors to neurodevelopmental disorders, including autism, schizophrenia, epilepsy, and intellectual disability [8,9,13,14]. These CNVs arise at hotspots that produce identical breakpoints across all carriers, eliminating the possibility of using breakpoint variation to map driver genes. A classic illustrating example of this is the 22q11.2 deletion, which affects 30–50 protein-coding genes and confers a dramatically increased risk for schizophrenia (odds ratio [OR] ~80)[15]. Because atypical 22q11.2 deletions are rare, occur in a gene-dense region, and produce largely pleiotropic effects, specific driver genes remain elusive, limiting translational potential; even integrative analyses that combine transcriptomic, functional, and common-variant data have been inconclusive, instead pointing to multigenic contributions [16–18]. Together, these observations underscore the ongoing complexity of identifying risk genes through large CNVs.
>
> **Somatic CNVs are a developing gene mapping strategy.** Cancer provides clear examples of risk gene discovery enabled by focal somatic CNVs, which often amplify or delete single genes [19]. These events are more frequent in somatic cell types and are typically smaller in size, making them well suited to recurrence-based analyses [20]. This contrast highlights a key translational insight: systematic discovery of somatic CNVs may improve resolution for localizing risk genes in complex disorders. Detecting focal somatic CNVs in brain tissue has been proposed as a frontier for identifying pathogenic drivers within multigenic germline CNVs implicated in disorders such as schizophrenia [21,22]. Integrating somatic profiling with germline CNV maps and functional assays provides a path to dissect risk genes embedded in highly penetrant, multigenic loci.

## 1.2 Small coding variants

Small coding variants, including single nucleotide substitutions and short indels within protein coding sequence, are among the most interpretable and widely studied sources of genetic risk. Unlike CNVs, they implicate a single gene with high confidence, enabling more precise mapping between genotype and phenotype. However, their extreme rarity, particularly in the case of *de novo* mutations, and the difficulty of functional interpretation present distinct challenges for gene mapping.

**Protein truncating variants are a rare, high-confidence impact class.** Variants that disrupt transcripts, such as frameshifts, stop-gains, and splice-site disruptions, are classified as protein-truncating variants (PTVs). The enrichment of these focal LoF variants has proven highly informative for identifying risk genes in many disorders when aggregated across large cohorts [23].

PTV-burden approaches have been particularly instrumental in identifying risk genes for severe psychiatric disorders, including autism, epilepsy, intellectual disability, and, to a lesser extent, schizophrenia [24–27]. Although PTVs provide direct insight into gene-level contributions, they account for only a small fraction of cases in most psychiatric cohorts—even among individuals with severe presentations. For example, large-scale autism sequencing studies have found PTVs in highly constrained genes in fewer than 5% of cases [25,26] (**Box 1**). Moreover, even among individuals with PTVs in high-confidence autism risk genes, clinical outcomes are highly variable [25,28]. Differences in severity, comorbidities, and whether an autism diagnosis is made at all have all been reported. Thus, PTVs capture only a small fraction of the broader genetic architecture of autism.

Despite their limitations, PTV-based gene mapping remains among the most immediately translatable strategies, particularly when conferring protection. Nearly two decades ago, PTVs in *PCSK9* were shown to markedly reduce LDL cholesterol and lower the risk of coronary artery disease, leading directly to the development of targeted *PCSK9* inhibitors [29]. Similar protective effects have been observed in *HSD17B13* in nonalcoholic fatty liver disease and are the subject of ongoing clinical trials [30]. In syndromic contexts where PTVs or dosage imbalances drive pathology, several FDA-approved gene therapies now target dosage restoration [31,32].

---

**Box 1: Considerations for gene constraint and isoforms when predicting variant impact.**

The deleteriousness of small variants is often predicted computationally given the scale of observed coding variation across genomes. More than 100 variant effect predictor (VEP) tools and databases currently exist, most relying on protein sequence homology, evolutionary conservation, or machine learning models trained on pathogenic variants and predicted RNA or protein effects [33,34]. These tools vary widely in accuracy, concordance, and sensitivity to context-dependent effects [33,34]. While some incorporate molecular traits such as gene expression or chromatin state, most remain anchored to annotations derived from monogenic disease databases, such as ClinVar. Nonetheless, VEPs facilitate the differentiation of variants based on predicted impact.

Risk gene mapping through the burden of rare, deleterious variants often incorporates measures of haploinsufficiency and evolutionary conservation to identify genomic regions under negative selection, typically summarized as gene-constraint scores. These scores are commonly estimated computationally using LOEUF (loss-of-function observed/expected upper bound fraction) [35], where lower scores reflect stronger purifying selection against loss-of-function (LoF) variants, and pLI (probability of loss-of-function intolerance) [36], which estimates gene haploinsufficiency. LoF variants in dosage-sensitive genes or regions under strong purifying selection are more likely to cause severe phenotypes.

Alternative splicing, the process by which different exon combinations generate multiple transcript isoforms from a single gene, occurs in ~95% of multi-exon genes. This widespread mechanism is especially frequent in the brain, where isoform diversity is pronounced and dynamically regulated during development [37]. As a result, alternative splicing is critical for interpreting the impact of any functional variation, yet it remains largely overlooked in risk gene mapping studies. The importance of isoform context has been experimentally validated in oncology, where the pathogenicity of *BRCA2* PTVs depends on whether the disrupted exon is incorporated into biologically relevant transcripts [38]. Cross-field in psychiatric genetics, isoform-aware interpretation remains underdeveloped despite the high prevalence of splicing in brain-expressed genes, but it is increasingly recognized as essential for accurate gene-level inference [37,39].

**Missense variants face interpretive ambiguity.** Missense variants, which alter amino acid identity, can be aggregated to implicate risk genes, though their pathogenicity is typically inferred computationally (**Box 1**). As an example of contemporary practice, psychiatric genetics often evaluates missense variants using composite metrics such as the MPC score, which combines predicted functional impact with regional intolerance to variation [40]. Rare variants predicted to be more pathogenic are enriched in case cohorts, particularly within genes under strong constraint, supporting their utility in discovering risk genes [24–26,41]. However, this enrichment is partly collinear with gene-level constraint, since the same features that elevate MPC pathogenicity score also mark genes depleted of missense variation. Thus, while MPC and other VEPs effectively prioritize pathogenic missense variants, dependence on constraint limits inference in other genes.

Beyond predicted pathogenicity, the functional interpretation of missense variants remains complex: (1) genes can harbor both GoF and LoF missense mutations with distinct phenotypic outcomes, as exemplified in *SCN2A* where GoF variants associate with epilepsy and LoF variants with autism [42]; (2) missense-enriched genes often overlap with those under strong PTV burden but can also be distinct, suggesting different mechanisms of pathogenicity; and (3) inherited missense variants show greater phenotypic variability than *de novo* ones [26], which cannot be fully determined without parental data. Despite integrative models that incorporate VEP-stratified missense variation alongside CNVs, PTVs, and inheritance mode, interpretation is still constrained by functional ambiguity and context dependence [43].

Addressing these interpretive challenges is critical, as missense variants play a foundational role in risk gene discovery. Indeed, the first disease gene identified in humans, *HBB* for sickle cell anemia, was discovered through a missense variant, where the abnormal hemoglobin protein was observed before the causal nucleotide change was mapped [44]. This case highlights how protein structure mediates variant effects—a principle that remains central as modern tools like AlphaFold enable structural prediction from sequence [45]. The sickle cell variant (rs334) illustrates these VEP limitations: though recessively pathogenic and phenotypically evident in carriers, its high frequency and context-dependent fitness yield inconsistent scores, with MPC in particular classifying it as benign. Together, these examples underscore the need to integrate molecular, phenotypic, and population context into gene-mapping frameworks, particularly for rare missense variants prone to misclassification by VEPs.

To overcome these limitations, experimental and structural strategies have emerged. Deep mutational scanning systematically assays nearly all possible missense changes in a gene, generating high-resolution functional maps, as demonstrated for *BRCA1* and *PTEN* [46]. In parallel, structural analyses situate observed mutations in three-dimensional protein space, revealing hotspots at catalytic or interaction interfaces that linear models cannot resolve [47,48]. Both approaches highlight how functionally convergent variants can cluster in ways obscured by sequence-level predictors. Extending these strategies to more genes could confidently isolate disruptive variants.

**Risk discovery on the X chromosome faces numerous challenges.** Male hemizygosity and the X-chromosome's reduced genetic diversity relative to autosomes confound metrics of its gene dosage sensitivity and constraint [49]. In females, phenotypic variability is further shaped by mosaic X-chromosome inactivation (XCI), a clonally maintained process that silences one of the X-chromosomes early in embryonic development. Although XCI can be quantified, tissue-specific mosaicism limits inference in relevant contexts, complicating interpretation even in all-female cohorts. Furthermore, an estimated 15–30% of genes escape XCI, leading to atypical expression [50]. Within the field of psychiatric genetics, this complexity is even further compounded by the female protective effect in autism and other neurodevelopmental disorders, where affected females carry greater mutational burdens than males [51,52], although it is not known if or how much of it is due to chromosome X. Together, sex-specific mechanisms and X-linked architecture pose persistent barriers to gene discovery.

To address these challenges, induced pluripotent stem cell (iPSC) models offer a promising experimental direction. Researchers have generated isogenic lines from female donors that differ only in which X chromosome is active [53]. This approach facilitates analysis of X-linked gene dosage effects, particularly in differentiated cell types, but is complicated by XCI erosion, in which the inactive X chromosome becomes partially reactivated [54]. Such models, while powerful, underscore the need for rigorous validation of XCI status and context-specific expression to ensure accurate interpretation of X-linked gene function.

**The rarity of recessive combinations limits insight.** Recessive inheritance can be viewed as a continuum shaped by haploinsufficiency and evolutionary constraint, with apparent recessivity emerging when heterozygous variants are tolerated but become pathogenic in combination. This model, encompassing both homozygous and compound heterozygous configurations has been central to identifying risk genes for Mendelian disorders. In contrast, detecting recessive and other non-linear contributions to polygenic disorders remains far more challenging, as the likelihood of two rare pathogenic variants affecting the same gene in a single individual is exceedingly low.

Two main strategies have emerged to overcome this combinatorial rarity. First, consanguineous pedigrees and extended families naturally enrich for recessive variants, enabling successful mapping of recessive risk genes with in disorders such as intellectual disability, cardiomyopathy, and other congenital conditions [55–57]. However, interpreting findings from these populations requires caution, as background variation, founder effects, and population-specific allele frequencies can obscure true pathogenicity or artificially inflate locus-specific recurrence (see **Box 3** for additional discussion).

Second, recessive effects can be revealed through autosomal hemizygous states, particularly when a coding variant occurs in trans with a large deletion. In 22q11.2 deletion syndrome, for instance, LoF variants in the hemizygous genes can unmask recessive effects, contributing to phenotypic variability among individuals carrying the same deletion [58,59]. As detection sensitivity improves for smaller CNVs, it may become increasingly possible to identify additional recessive risk genes without requiring large, syndromic deletions or consanguineous pedigrees.

*1.3 Atypical variant classes*

Beyond small coding variants, more complex variants classes remain underrepresented in risk gene mapping. These variant classes require specialized detection tools that require awareness of other proximal variation and their haplotype arrangements. Here we briefly overview their utility.

**Tandem repeats are an impactful but poorly resolved class of genetic variation.** Tandem repeats (adjacent motifs of 2–20 bp) are a distinct class of highly mutable genetic variation that forms through repeat-specific mechanisms. Rare expansions of tandem repeats are causal in over 50 genetic disorders, including Huntington's disease and Fragile-X syndrome [60]. Beyond these pathogenic expansions, tandem repeats are increasingly studied in polygenic disorders, both as potential causal elements and as common polymorphic loci offering fine mapping resolution due to their high mutation rates [61,62]. While identifiable in short-read WGS using various tools [63], caution is warranted when applying tandem repeats to risk gene mapping: (1) Their genomic catalog remains incomplete, with substantial disagreement across reference panels [64,65]; (2) High mutability creates population-specific allele distributions that complicate cross-population analyses; (3) Pathogenicity thresholds are often non-linear, with effects emerging only beyond certain repeat lengths; and (4) Short-read resolution is still insufficient for clinical accuracy [63]. Despite these limitations, tandem repeats contribute meaningfully to phenotypic variation and offer a unique opportunity for risk gene prioritization.

**In-frame indels and clustered substitutions are overlooked genetic variation.** Unlike missense variants, which alter amino acid identity, or frameshifts, which disrupt the entire transcript, in-frame indels add or remove amino acids without shifting the reading frame. In-frame indels contribute to Mendelian disorders, exemplified by the recessively inherited F508del mutation in *CFTR*, a LoF in-frame deletion that causes the majority of cystic fibrosis cases [66]. This variant illustrates the context-dependent challenge shared by in-frame indels and SNVs: *CFTR* is tolerant of LoF mutations and F508del has a high carrier frequency (~4% in Northern Europeans), which would drive VEPs to underprioritize its pathogenicity.

More accurate functional prediction, especially for in-frame indels of non-Mendelian impact, requires models that emphasize protein folding dynamics or domain-level perturbation. Functional screens and deep mutational profiling, which capitalize on in-frame indels as common artifacts in CRISPR, show that these variants span a spectrum of effects, from benign to LoF and, less commonly, GoF [67,68]. Yet, despite their functional weight and preservation of reading frame, in-frame indels remain underutilized in gene discovery efforts.

In addition to in-frame indels, multi-nucleotide variants (MNVs), which involve two or more nucleotide changes on the same haplotype and often within the same codon, are also frequently misinterpreted. MNVs can arise through specific mutagenic mechanisms, particularly in cancer, and comprise approximately 2% of *de novo* variants [69]. Critically, MNVs can result in amino acid substitutions that are only possible through multiple simultaneous nucleotide changes. Estimates from whole-exome sequencing suggest that approximately 60% of MNVs produce amino acid substitutions distinct from those caused by either variant alone, including PTVs that are missed when variants are unphased [70]. To address this, haplotype-aware annotation tools have been developed [71,72] and large-scale resources such as gnomAD implement read-based phasing to distinguish *cis* from *trans* configurations of MNVs [70]. Both in-frame indels and MNVs thus require standardized post-calling refinement to distinguish these variants from traditional SNVs and reduce false signals, especially in rare variant analyses where even a single misclassification can mislead gene discovery.

*1.4. Additional complexities in classifying variant impact*

In addition to classical models of variant penetrance, such as dominance, recessiveness, and variable expressivity, other factors influence how genetic variation contributes to phenotype. These include background genetic variation, discussed in later sections, and context-dependent mechanisms that modulate variant impact. In this section, we will briefly outline key examples.

**Parent-of-origin effects mask variant impact.** Genomic imprinting is a well-characterized epigenetic mechanism in which gene expression is selectively silenced based on the parent of origin. As a result, deletions on the expressed homolog—or, less commonly, uniparental disomy or imprinting defects—can lead to complete loss of gene function due to epigenetic silencing of the other copy. While the set of imprinted genes in the human genome is established, recent evidence indicates that imprinting can vary across tissues and between individuals, suggesting a more dynamic and context-dependent regulatory landscape [73,74]. Parent-of-origin effects have been particularly important in developmental biology, cancer, and psychiatric genetics [75–77], but they remain largely unaccounted for in most risk gene mapping studies. Incorporating imprinting-aware analyses may help unmask functional variation that would otherwise be obscured in standard association frameworks. While a few statistical frameworks and tools have been developed to model parent-of-origin effects (e.g., EMIM[78] and POIROT[79]), these methods are not yet widely adopted. Most large-scale gene-mapping studies do not incorporate imprinting or parent-specific effects in their standard analyses.

**Somatic mosaicism is limited by tissue type and technical noise.** Mutations that arise in specific cell lineages can influence phenotype but are often missed when genetic material is obtained from unrelated tissues. Detecting mosaic mutations typically requires sequencing the affected tissue, which is not always feasible in living donors, or identifying low-frequency variants, which has traditionally depended on ultra-deep sequencing to distinguish true signal from technical noise [80]. Advances in sequencing technologies, including long-read platforms and the higher accuracy sequencing-by-binding approach [81], are positioned to reduce depth requirements and improve detection of somatic mosaicism.

**Variant impact can differ significantly by biological sex**. *BRCA2* mutations classically exemplify sex-specific risk differences, as pathogenic variants differentially impact male and female cancer risk. In neurodevelopmental disorders, the "female protective effect" describes a separate phenomenon where females carry a higher burden of deleterious variants or show milder phenotypes than males with the same mutation [82]. For instance, PTVs in *CHD8* are more commonly observed in males with autism, and mouse models support this observation, showing that *CHD8* loss reduces neurotransmission in males but not females [83]. These examples demonstrate how sex bias is critical for interpreting variant effects, particularly in complex, polygenic disorders. Sex specificity also has implications for VEPs and related databases, where male-biased sampling, as is the case in many autism studies, may skew pathogenicity assessments.

## 2. Intermediate variation and effect sizes

Current methods for identifying risk genes are optimized for either common or rare variants. As described in our introduction, variants in the intermediate-frequency range (0.1–5%), particularly those with intermediate effect sizes (e.g., OR 1–2), are systematically under-detected by both approaches.

Statistical power is the strongest factor limiting discovery in this intermediate frequency band. While biobank-scale GWASs of continuous traits (e.g., height, molecular phenotypes) can achieve power for variants with MAFs down to 1%, associations below 1% remain inaccessible under conventional designs, particularly in case-control contexts. For instance, detecting a variant at MAF = 0.1% with OR = 1.1 at genome-wide significance ($\alpha = 5 \times 10^{-8}$) would require over 5 million cases and an equal number of controls. Burden-based methods for rare variants face parallel limitations: under reasonable assumptions (e.g., 20 variants per gene, 0.1% MAF, 50% causal, exome-wide $\alpha = 2.5 \times 10^{-6}$), a SKAT-O test would still require >500,000 cases to achieve 80% power. These power limitations explain the aforementioned enrichment of benign-labeled variation in this intermediate range (**Fig 1**).

Risk gene discovery within the intermediate frequency and effect size spectra is further hindered by methodological dependencies on gene constraint (**Box 1**). Current gene-mapping and variant prioritization tools often rely on metrics such as gene-level intolerance to variation or variant-level conservation, prioritizing highly deleterious, large-effect loci. By contrast, intermediate effect size variants and risk genes may be weakly selected, structurally tolerated, haplosufficient, and absent from curated pathogenic databases. These features lead VEPs to misclassify them as benign and reduce power in burden-based tests, especially when stratification by predicted deleteriousness is required for signal.

Despite these challenges, variants in the 0.1–5% frequency range and risk genes with lower effect sizes are predicted to represent a substantial source of heritable variation in human traits and diseases [3,84]. The following sections examine emerging approaches to detect these variants: joint gene analyses that capture combinatorial effects, phenotype refinement strategies that increase statistical power, and network-based methods that leverage biological relationships (**Fig 2**).

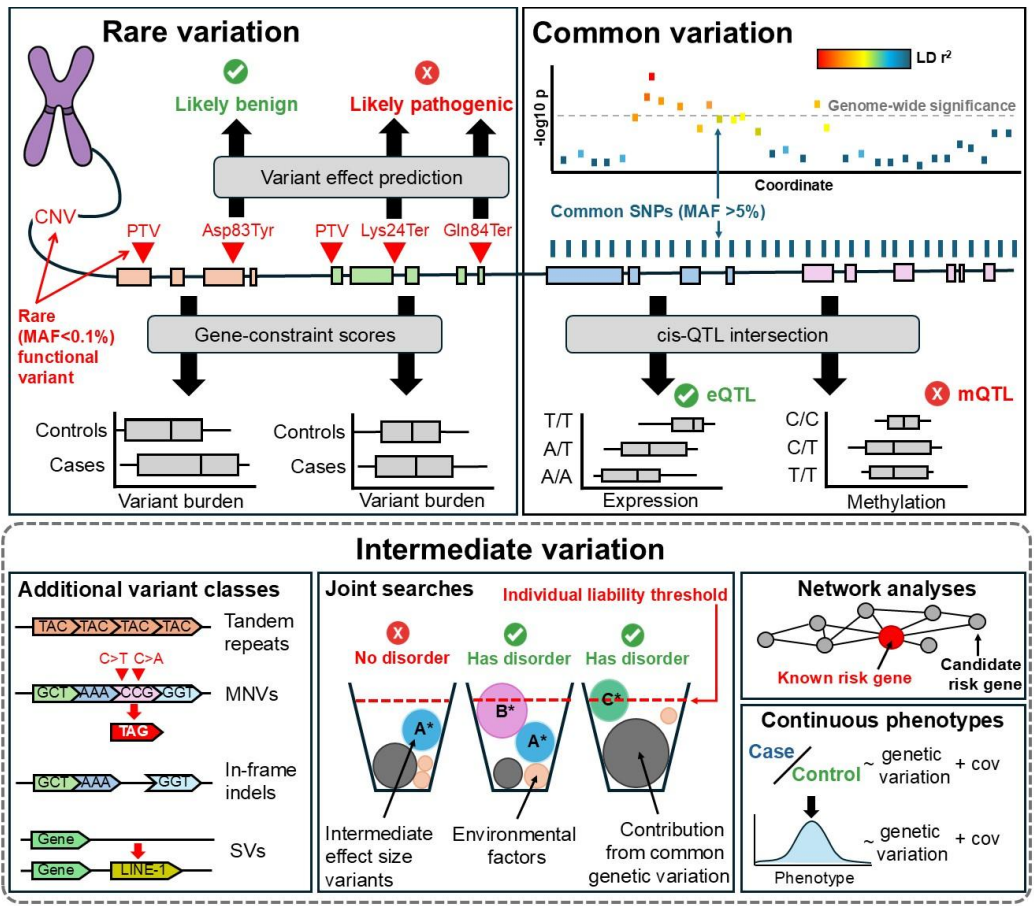

**Figure 2. Approaches to risk gene discovery across rare, intermediate, and common variation.** The threshold model for individual liability states that a disorder manifests once combined genetic and environmental risk surpasses a critical threshold. Liability reflects the cumulative effects of many variants and exposures, with diagnosis occurring after this threshold is crossed. Rare functional variants can individually reach liability, conferring strong discovery power in highly constrained genes. Common variants, by contrast, act collectively to define genomic intervals linked to disorder risk and contribute incrementally to liability. Methods for detecting intermediate-frequency variants remain underdeveloped. Expanding variant classes, integrating joint searches across multiple functional mutations (e.g., A*, B*, C* in distinct genes), incorporating common variation and environmental factors, and applying network and quantitative trait analyses all increase discovery power.

## 2.1 Joint searches

Testing for enrichment of multiple intermediate effect size variants within the same individual, or combinations of these variants with high polygenic burden measured from common variants, offers an alternative discovery approach that aggregates their collective signal, particularly in affected individuals who lack high-impact variants. This subsection focuses on mapping intermediate effect size risk genes through paired-gene analyses, the simplest combinatorial unit beyond single-gene models. While higher-order combinations almost certainly contribute to disease risk, their identification faces severe multiple-testing burdens. We review observational and experimental approaches for detecting joint effects on liability and distinguishing independent from dependent gene pairings (i.e., additive from synergistic; **Box 2**). We also consider conceptual extensions beyond gene pairs, such as the joint influence of rare variants in one gene and polygenic background, where genome-wide common variant burden may modulate penetrance or obscure locus-specific effects.

---

**Box 2: Interactions between genetic risk factors.**

Under an assumption of genetic independence, two loci may influence individual liability through additive or synergistic interactions. In additive models, each locus contributes separately to liability, with joint effects approximating the product of individual risks (see illustration right). For example, under a threshold liability model, LoF mutations in any two genes conferring independent sub-threshold impacts* may only cause disease when both are present in the same individual.

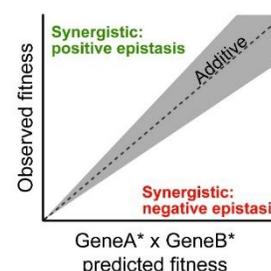

In contrast, synergistic pairs confer risk only in specific combination, with liability emerging from the identity of impacted genes. Such relationships can produce outcomes that depart from expected outcomes in magnitude or direction. This framework reflects classical epistasis, where negative interactions (e.g., synthetic lethality) exacerbate liability, while positive interactions mitigate it.

---

**Conditional risk may be assessed from jointly disrupted gene pairs**. Cancer provides a particularly amenable context for identifying co-occurring driver genes, due to high somatic mutation rates and the selective pressure of clonal expansion. In tumors with frequent somatic alterations, studies often evaluate gene pairs using models like hazard ratios to assess prognostic impact [85]. These analyses reveal conditionally pathogenic relationships, where the effect of one mutation depends on the presence of another, and survival outcomes vary by combination. A canonical example arises in colorectal cancer, where mutations in *APC*, *KRAS*, and *TP53* frequently co-occur [85,86]. None of these is sufficient alone for full transformation, underscoring the requirement for multiple driver events. Temporal ordering also plays a role: *APC* loss typically initiates tumorigenesis, with later mutations (e.g., in *KRAS*, *TP53*) driving progression [86,87]. Paired analyses thus clarify both cooperative interactions and sequential acquisition, features often obscured in single-gene models. Clonal selection enriches these functional combinations within tumors, enabling detection of gene pairs that would otherwise remain too rare to identify.

In germline variation, the absence of clonal enrichment makes detecting conditionally pathogenic pairs far more challenging. For example, neurodevelopmental disorders have lower variant frequencies and complex trait architectures, but their high polygenicity and substantial heritability create the potential for signal. In this context, pathogenicity may be inferred from a statistical excess of jointly disrupted gene pairs across large cohorts. To our knowledge, only one study has applied a combination of transmission-based candidate selection with paired-gene burden testing to map risk genes [88]. This analysis identified 50 novel gene pairs for autism in which rare, inherited deleterious variants confer risk only when both genes are disrupted. These pairs show distinct characteristics: lower evolutionary conservation and enrichment for mild psychiatric traits even among single-gene carriers. Despite this progress, paired-gene frameworks remain underutilized, constrained by sample sizes and incomplete variant ascertainment.

**Experimental systems test interactive effects**. Computational approaches for modeling interactions among multiple functional variants within the same genome, such as LoF alleles in different genes, remain limited. Progress has been hindered by the scarcity of empirical observations, the combinatorial complexity of potential interactions, and the lack of statistical frameworks capable of capturing these effects with sufficient power.

In yeast, array-based platforms have enabled the systematic construction and high-throughput fitness profiling of double mutants [89]. Crucially, these experiments empirically distinguish additive from synergistic effects: under an additive model, the combined phenotype of two individual gene knockouts follows the multiplicative product of their individual fitness values, typically measured by colony growth (**Box 2**). Synergistic interactions are inferred from significant deviations from this expectation. Large-scale screens show that fewer than 5% of double mutants display such synergistic deviations, highlighting that most genes paired at random behave additively under standard growth conditions [90,91]. This predominance of additivity simplifies joint searches because we can assume most paired effects sum rather than requiring specific combinations.

In humans, CRISPR–Cas9 editing has enabled double-mutant screening, though applications have largely focused on cancer-related genes [92,93]. While limited in scope, these studies similarly report that fewer than 1% of tested gene pairs show significant deviations on fitness from additive expectations. The strength and direction of interactions are often cell-type specific and, as observed in yeast, dependent on pathway context. Redundancy across genes within the same pathway is hypothesized to buffer gene loss as LoF mutations in multiple steps of the same pathway more often result in negative interactions. However, as is observed in colorectal cancer, dependent pairs may still be enriched within disorders precisely because of these interactions.

These experimental findings inform strategies for detecting gene pairs in human germline variation. Quantitative phenotypes for disorders, such as social responsiveness scores in autism, could be used to quantify interactive effects of risk gene combinations. Assuming additivity as the baseline, as supported by yeast and cancer models, deviations in phenotypic scores among individuals with variants in two or more genes relative to the product of single-gene effects would indicate synergistic interactions in complex disorders.

**Joint analysis of rare and common variation could enhance risk gene discovery.** Polygenic risk scores (PRSs), typically derived from GWAS, summarize the cumulative contribution of common SNPs (see Section 3 for further discussion of common variation). In many disorders, individuals with the highest PRSs can exhibit effect sizes comparable to those conferred by single large effect size variants. For example, in schizophrenia, the odds ratio between individuals in the top and bottom 1% of the PRS distribution is comparable to that of specific pathogenic mutations [15]. Accordingly, PRS serves as an individual-level statistic whose effect magnitude is captured by score percentile.

While PRSs can substantially influence individual liability, they are rarely sufficient on their own to exceed diagnostic thresholds. In contrast, high-impact variants, particularly *de novo* mutations, often confer risk independently of background polygenic burden due to their large individual effects [94,95]. However, PRS can still modulate phenotypic expression among carriers, influencing severity, age of onset, or comorbid features [96–98].

Incorporating PRS into gene discovery frameworks may help reveal intermediate effect size risk genes whose contributions are obscured by background polygenic variation. Though established methods such as CORAC, RICE, and CLIN_SKAT integrate rare and common variant signals to improve risk prediction and uncover shared mechanisms [99–101], they do not extend this to gene-mapping approaches. To our knowledge, no current approaches explicitly leverage PRS to identify novel risk genes whose penetrance may be modulated by polygenic background.

## 2.2. Phenotype-focused approaches

A central challenge in risk gene mapping is the dichotomization of case and control. This framework requires genetic variants with effects large enough to associate with reaching a diagnostic threshold, such as a clinical autism phenotype or a tumor classified as malignant. However, most traits exist on a spectrum. In autism, this spectrum is reflected in the continuous distribution of traits like social responsiveness in the general population, with diagnosis marking individuals who exceed a pathologized percentile. In cancer, precancerous changes accumulate over decades, with malignancy representing the end point of continuous progression. Genetic variants that shift social responsiveness in the general population or accelerate adenoma formation are biologically and clinically meaningful yet remain undetectable in classic case-control studies. This section reviews how phenotype refinement creates opportunities to detect intermediate impacts and novel risk genes.

**Refinement of outcomes improves the power of association analyses.** Traditional PheWAS, which tests pre-selected variants against multiple phenotypes, characterizes pleiotropic effects of known variants but are not designed to discover new risk genes [102]. For example, recent PheWASs tested nominally autism-associated variants for links to quantitative traits, identifying genes tied to specific phenotypic dimensions [103,104]. This approach is valuable for understanding within-case heterogeneity though limited to pre-selected variants. In contrast, discovery-oriented approaches, including analyses that test all variants against multiple phenotypes, can identify novel risk genes but are still limited by outcome dichotomization.

Quantitative trait-based strategies confer a statistical advantage which can augment discovery in the intermediate frequency range. In an association analyses of over 250,000 individuals from the UK Biobank, where all detectable variants were tested against 17,000 binary and 1,400 quantitative phenotypes[105], variant-level associations were nearly eight times more common for quantitative traits, and gene-level associations were identified in 54% of quantitative phenotypes compared to just 5.4% of binary traits. Such gains in detection power emphasize the utility of continuous traits for uncovering lower effect size loci that are missed in binary analyses.

A similar quantitative trait-centered logic has advanced aging research. Rather than focusing solely on mortality, a survival-based late-stage outcome that may occur decades after the onset of biological changes, researchers increasingly use composite measures of physical function, chronic disease burden, and molecular traits such as cellular senescence and DNA methylation to quantify aging earlier in the lifespan. Refining aging phenotypes in this way has enabled the discovery of risk genes that are not associated with mortality-based endpoints. For example, a GWAS of healthspan, defined as survival without major age-related disease, identified several loci not linked to mortality or to any single disorder [106]. Similarly, a GWAS of methylation-derived age acceleration revealed genetic associations not captured by survival-based endpoints [107]. This refinement has informed drug development, including senolytic compounds targeting cellular senescence [108] and metformin trials guided by epigenetic aging biomarkers [109].

**Within-case genetic mapping probes heterogeneity to identify subtype-specific risk factors.** Complex polygenic disorders display phenotypic heterogeneity, both in symptoms and onset timing. As such, comparing individuals with the disorder can identify genetic components that may be missed in standard case-control designs. Alzheimer's disease illustrates this, with early-onset cases showing distinct genetic risk from the more common late-onset form [110]. Similarly, comparisons of amyotrophic lateral sclerosis cases identified a novel gene and environmental correlations specific to early-onset subtypes [111]. Within-case comparisons have also helped clarify how biomarkers relate to disorder states as applied in studies of chronic kidney disease [112]. By isolating within-disorder variation, these analyses reveal how genetic effects are limited to certain biological states or exposures, providing insight into the conditional architecture of disease risk.

Heterogeneity among cases can also reflect sex-specific effects or population-specific genetic variation, the latter of which will be addressed in later sections on common variation. With respect to sex, its role in shaping genetic risk is well established in case-control studies, as seen in phenomena like the aforementioned female protective effect in autism. However, partitioning cases by sex in within-case analyses requires caution and may not be appropriate. Sex-specific variation may appear enriched not because of sex-specific risk genes, but because of sex-specific penetrance. As a result, within-case gene mapping can be highly confounded by these secondary effects unrelated to variant function, including hormonal, developmental, or environmental modifiers.

**Molecular phenotypes provide another avenue for phenotype refinement.** Rare and intermediate frequency variants are typically underpowered in conventional analyses of gene expression, splicing or methylation, yet can produce pronounced outlier signals in these molecular phenotypes. These outliers reveal functional effects of variants that lack statistical power in case-control studies but measurably alter molecular traits. Methods such as Watershed exploit such signals to predict the regulatory effects of other variants by identifying individuals with extreme expression patterns driven by otherwise cryptic genetic variation [113,114]. This framework encompasses both coding and non-coding variants that impact transcription or splicing in ways not captured by protein-centric annotations. Extensions of Watershed could incorporate more permissive outlier thresholds or weight intermediate deviations, yielding a continuous measure of regulatory impact. Such refinements would allow variant prioritization in risk gene mapping to focus on alleles that influence gene regulation rather than solely those causing LoF effects. Developing more precise estimates of variant consequences for gene function remains a critical frontier for discovering novel genetic risk factors.

*2.3. Network analyses*

Incorporating external biological data can prioritize genes or variants lacking strong marginal signals by leveraging their relationship to known risk genes. Such approaches are well suited to nominate lower-impact risk genes and regulatory variants missed by traditional designs, and to improve variant assessment.

**Network integration refines candidate genes.** In polygenic disorders, risk genes often cluster within regulatory or co-expression networks, motivating approaches such as DAWN that leverage these relationships to prioritize candidates through co-expression with known risk genes [115]. As DAWN's developers note, co-expression or interaction alone does not prove regulatory function, thus network evidence is most powerful when combined with genetic data. Network-based methods are also limited by the need for relevant tissue and developmental context. *CHD8*, a major risk gene in several developmental disorders, illustrates the complexity of network interpretation: it is expressed broadly across cell types and stages and interacts with over 100 proteins in addition to DNA, yielding an extensive set of network-derived candidates that require further validation to distinguish functional from incidental connections.

More recent methods, such as VBASS[116] and DYNATE[117], integrate cell-type-specific expression data with mutational burden to refine associations. VBASS uses Bayesian modeling to weight variant burden by expression levels in relevant cell types, while DYNATE dynamically aggregates genomic regions to identify enrichment of disease-associated variants. This allows functional context to inform candidate risk gene selection while preserving the requirement for genetic evidence. Notably, VBASS is designed exclusively for *de novo* variation, which limits its applicability to inherited variants. DYNATE can handle rare variants more broadly but still faces challenges with inherited variants that have weaker individual effects. For such variants, network evidence may need to be supplemented with family-based transmission analyses or functional validation approaches.

## 3. Common variation and low effect sizes

Common variants (MAF > 5%) account for 12.8% of SNVs in 1kGP, including 5.9% of missense variants and 4.2% with predicted protein impact (**Fig 1B**). In GWAS, common variants act as proxies for nearby causal alleles through their linkage disequilibrium (LD) but individually confer minimal effect on liability (**Fig 2**). Such signals can implicate genomic regions yet rarely pinpoint causal variants, which may instead be rare, regulatory, or distributed across multiple sites. Resolving this architecture requires fine-mapping and complementary strategies. Accordingly, our review of common variation focuses not on GWAS methodology itself, but on emerging approaches in fine-mapping, ancestry-aware analyses, and integrative or experimental methods that aim to close this gap and move association signals toward causal gene inference.

### 3.1. Fine-mapping within LD blocks

While GWAS highlights regions of association, pinpointing the causal variants and corresponding risk genes remains a major challenge. Fine-mapping addresses this by prioritizing variants within genome-wide significant loci that are most likely causal (or most proximal), using summary statistics and LD patterns [118]. Most approaches first estimate the number and identity of causal variants by testing models defined by local association signals, LD, and a user-specified maximum causal variants to test per locus. Locus independence is then assessed by conditioning on the putative causal variants.

> **Fine-mapping is sensitive to LD structure.** Fine-mapping has increased the number of predicted causal variants in many GWASs, but performance is highly sensitive to local LD patterns and allele-frequency spectra that vary across populations. For this reason, GWAS and fine-mapping efforts have largely been performed within populations with similar ancestral backgrounds, particularly European. To study ancestrally-diverse cohorts, tools such as SuSiEx enable cross-population fine-mapping by incorporating ancestry-specific LD patterns and allele frequencies [119]. Applied to breast cancer, these approaches improve resolution and expand the catalog of risk genes and loci [120].

> Demographic history plays a substantial role in shaping LD and allele frequencies (**Box 3**). In addition to greater SNP density, African LD blocks are notably smaller, reflecting fewer historical bottlenecks and higher within- and between-population diversity, including variation at *PRDM9* which influences meiotic crossovers [3,121]. For this reason, fine-mapping resolution improves when comparing only African populations [122]. In use, a GWAS of prostate cancer in African men identified four novel loci, six signals at loci exclusively polymorphic in African populations, and African-specific risk haplotypes with singularly strong effect sizes [123]. Methodological advances and expanding, diverse sequencing cohorts are improving the prospects for risk-gene mapping from common SNPs.

**Box 3: Timescale constraints on LD and variant age.**

Fine-mapping resolution depends not only on cohort size and LD structure but also on the evolutionary timescale over which variants have segregated. Newly arisen variants first appear on haplotypes that span entire chromosomes. Recombination gradually breaks down this linkage, narrowing the region of correlation around the causal site. Under standard models of LD decay [124] and standard sex-averaged human recombination rate (~1.2cM/Mb), it takes approximately 200 generations for the squared correlation ($r^2$) between a causal variant and a marker 300 kilobases away to drop below 0.1, the conventional threshold for statistical independence (see illustration right). Reaching the same level of independence within 60 kilobases, a distance comparable to the median human gene length, requires approximately 1,700 generations.

Common variants identified through GWAS therefore reflect tagged alleles that have persisted for hundreds to thousands of generations without removal by purifying selection. Their long-standing presence indicates a history of sufficient tolerability within the relevant demographic and selective context.

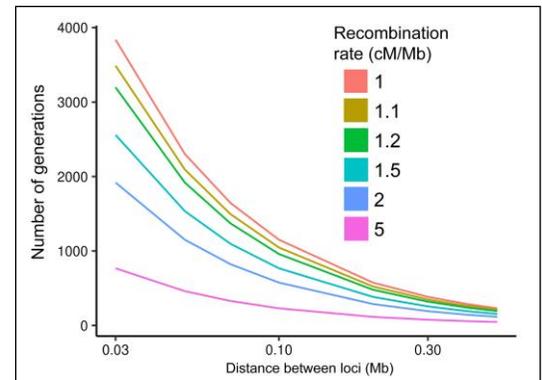

LD decay with genetic distance is approximated by the exponential model $D_t \approx D_0 e^{-\theta t}$, where $D_t$ is the disequilibrium at generation t (here 0.1), $D_0$ is the initial LD (here 1), and θ is the sex-averaged recombination frequency (in cM, converted to Mb using the recombination rates). These timelines are further extended in the presence of assortative mating or population structure.

Admixture and ancestry also introduce confounding in the interpretation of rare variants. Without trio data, phasing novel variants is typically performed by imputation, which is unreliable for rare alleles. In such cases, haplotype identity remains ambiguous, and the ancestry of the variant cannot be resolved with confidence. Furthermore, variant age is similarly dependent on ancestral background. An analysis of $f_2$ variants in the 1kGP, defined as variants observed exactly twice in the cohort due to inheritance from a shared ancestor (MAF ~0.2%), demonstrated that $f_2$ variants found within African populations are substantially older, while those in admixed American populations are younger by at least an order of magnitude in generational time [125]. Variants matched for frequency may therefore differ dramatically in their age, transmission history, and informativeness for fine-mapping or inference of selective pressures.

**Disagreement between rare- and common-variant findings endures.** Recent large-scale fine-mapping studies underscore the discrepancy between risk genes identified by rare versus common variation. For example, in the most recent schizophrenia GWAS with over 75,000 cases, only 5 of 32 genes implicated through rare-variant burden analyses overlapped with the 2,195 genes proximal to the 287 significant loci [24,126]. This divergence is expected: rare alleles are more likely to produce severe phenotypes and, when distributed across multiple sites within the same gene, dilute linkage signals. In breast cancer, the contrast is equally stark: *BRCA2* contains over 5,000 pathogenic variants in ClinVar, yet the most common has a gnomAD allele frequency of just 0.02%. This diffuse allelic diversity explains why *BRCA2* does not yield reliable GWAS signals. These discrepancies further reinforce that GWAS implicates a distinct class of risk genes, where subtle regulatory or functional effects are driven by common alleles with extensive transmission histories (**Box 3**). As explored in a later section, eQTL colocalization remains a leading approach to interpret regulatory potential, consistent with the idea that common alleles are more likely to modulate risk gene expression.

**Association mapping in admixed individuals is increasingly feasible.** Ancestry confounding is further complicated by admixture, as local ancestry can diverge from genome-wide estimates. This challenges the standard use of ancestry PCs generated from genome-wide genetic variation as a covariate in GWAS. For example, if a causal variant lies on a European haplotype in an admixed individual with only 10% overall European ancestry, the correct reference should be European LD and allele frequencies on that haplotype. Moreover, in admixed cohorts, multiple causal variants within the same risk gene harbored on ancestrally diverged haplotypes can generate conflicting or null associations in addition to their different transmission histories (**Box 3**). For these reasons, admixed individuals are routinely removed from GWAS.

Methods such as Tractor have begun to overcome GWAS in admixed populations with local ancestry-aware regression models, which have improved risk gene discovery and resolution in African American populations [127]. Local ancestry is typically inferred with tools such as FLARE or SHAPEIT5, which use phased genotypes and reference panels to assign ancestry at each locus, allowing inference across multiple ancestries [128,129]. Of note, most methods are tested on and designed around admixture across only two superpopulations (e.g., European and African ancestry). GWAS in admixed populations with three or more ancestries necessitates larger sample sizes.

**Underutilized classes of common variation can reveal new associations.** GWASs almost exclusively use SNPs and small indels as exposures, yet other forms of common genetic variation offer distinct advantages. Tandem repeats, as discussed above in Section 1.3, mutate far more rapidly than SNPs and generate multiple alleles per locus, which can increase power and fine-mapping resolution. While PheWASs and *cis*-QTL studies of tandem repeats have been recently performed [130,131], dedicated methods for tandem repeat GWAS remain underdeveloped. Nevertheless, analyses of polymorphic tandem repeats in 1kGP show that their ancestry composition mirrors that of SNPs, indicating comparable ability to capture population structure and LD (**Fig 3**).

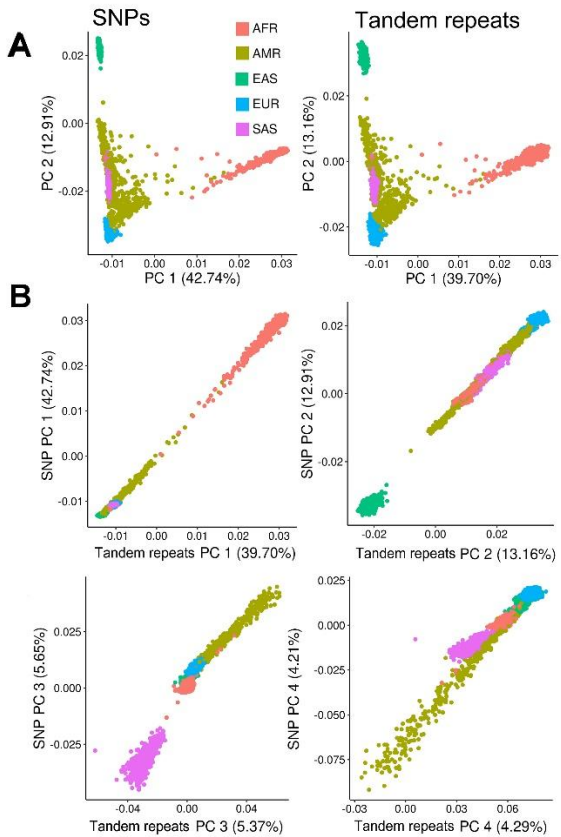

**Figure 3. Comparison of principle components from SNPs and tandem repeats. (A)** Using the 3,202 individuals from the 1kGP cohort, we genotyped tandem repeats at 380,935 autosomal loci from Ziaei Jam et al.[64] reference panel with ExpansionHunter [132]. The Ziaei Jam panel was filtered to motif lengths between 2 and 20bp and loci within protein coding sequence, introns, or 2.5kb immediately up or downstream of a protein coding gene. We calculated 20 principle components (PCs) from either tandem repeat genotypes or SNPs constrained to the same genic intervals. **(B)** Comparisons of the same PCs calculated from SNPs or tandem repeats as in panel A. Components are nearly identical for the first two PCs and explain a larger percentage of variance.

Structural variants (SVs) extend beyond CNVs to include inversions, complex rearrangements, and mobile elements, many of which segregate at common allele frequencies. Population-scale short-read studies can reliably detect only large SVs (>10Kb), which are typically rare [133]. Detection of smaller CNVs has improved with short-read sequencing and tools such as GATK-gCNV [133] and rCNV[134]. However, short reads often lack the precision needed to resolve variant boundaries or classify structural types, prompting the development of specialized algorithms and long-read sequencing efforts [135]. As a result, small CNVs remain understudied in most disorders but represent a promising direction for future gene discovery. Indeed, recent long-read sequencing of 1kGP samples has revealed abundant common SVs, with population frequency inversely proportional to variant size[135]. As sequencing and genotyping sensitivity improve, association mapping will increasingly be able to incorporate tandem repeats, SVs, and other variant classes—each with distinct mutation rates and LD structures—into risk gene discovery.

### 3.2. Aggregative strategies for gene discovery

Because individual common variants typically exert small effects through predicted regulatory capacity, integrating molecular readouts enables gene-level resolution beyond the simple "nearest gene" metric. This can be achieved through statistical overlap with regulatory QTLs, integrative models that predict expression, or experimental assays that directly test variant function. Together, these approaches shift common variant studies from proxy associations toward causal risk gene inference.

**cis-QTL colocalization improves gene prioritization.** *cis*-acting QTLs provide functional anchoring for GWAS loci by linking variants to gene-level changes in expression, splicing, or methylation. Colocalization between trait-associated variants and *cis*-QTLs supports the inference that regulatory perturbation of a given gene mediates disease risk. These overlaps are especially valuable because they can be tissue- and stage-specific, providing biological insight beyond statistical association. However, their utility is limited by QTL sample size, incomplete coverage of tissues and developmental contexts, and strong ancestry bias in available reference cohorts [136]. This limitation and potential for resolution is highlighted by recent work mapping expression and splicing QTLs in East African populations which identified extensive population-specific QTLs and improved fine-mapping resolution [137].

Transcriptome-wide association studies further integrate cis-eQTL effects into GWAS to test whether genetically predicted gene expression is associated with a trait, while colocalization approaches like coloc and eCAVIAR more directly evaluate whether the same causal variant underlies both molecular and disease associations [138–140]. These approaches can aggregate QTL information across variants, model uncertainty in causal assignment, and integrate tissue-specific expression to prioritize candidate genes. Yet, as with direct QTL overlap, their resolution depends on the quality and diversity of underlying QTL panels which are positioned to improve with increased multi-omic data generation.

**Experimental strategies assay common variant function at scale.** Massively parallel reporter assays (MPRAs) provide a powerful experimental approach to functionally validate common variants implicated by GWAS. A recent study applied MPRA to systematically test over 1,200 common and low-frequency variants (MAF > 0.5%) from non–small cell lung cancer GWAS loci by cloning each allele into reporter constructs and measuring their effects on gene expression in lung cancer cell lines [141]. The assay identified 82 functional regulatory variants and pinpointed 30 likely causal alleles, demonstrating how MPRAs can resolve LD-rich regions to link regulatory variants to genes. Nonetheless, MPRAs are labor-intensive and constrained by cell-type specificity and developmental context, posing particular challenges for their application to other complex disorders.

## Discussion

While cohort sizes and computational resources have grown exponentially, the pace of linking specific genes to human traits and diseases has decelerated. We demonstrate that this plateau does not indicate the depletion of biological insight but rather the inherent limits of prevailing statistical frameworks. Contemporary approaches remain most powerful at the frequency extremes of rarity and commonality, leaving the intermediate zone comparatively underexplored. This review highlights the limitations of these contemporary approaches and recent innovations to overcome them, particularly within the intermediate frequency band. This review's perspective is ultimately optimistic, highlighting opportunities to further push discovery efforts.

Variants in the intermediate frequency range remain underrepresented not because they are biologically trivial, but because their effects fall below the detection thresholds of prevailing models. Consequently, the relative depletion of pathogenic genetic variants in this range suggests their roles have yet to be identified. Developing methods to study the intermediate range have extensive translational purpose both as therapeutic targets and in understanding the full genetic architecture of disorders. Specifically, intermediate-frequency alleles are more likely than ultra-rare variants to segregate within families and populations, providing insight into the heritable architecture that underlies many disorders. This frequency range also forms the middle ground between Mendelian and polygenic paradigms which often produce distinct signals. Moreover, variants and genes that impart lower effect sizes may experience weaker selective pressure or greater modulation by environmental factors, presenting sources of phenotypic heterogeneity within risk architecture itself. Although current methods remain limited in quantifying these effects, the emerging approaches we described offer promising avenues.

Incomplete penetrance remains an ongoing limitation in interpreting variant effects across all classes, warranting specific discussion. It is often unclear whether a variant or gene has a genuinely modest effect on liability or a stronger effect that is not consistently expressed. This distinction is important: effect size reflects a variant's intrinsic impact, whereas penetrance reflects modulation by background genetic, environmental, and demographic factors. Penetrance is typically estimated as the proportion of variant carriers who exhibit the associated phenotype, but this approach is analogous to single-gene burden testing and thus the quantification of impact size. An alternative has recently emerged by reframing the question: given a variant or genetic profile, why does disease not occur? Using large-scale EHR-linked biobank data, recent work has applied machine learning to predict expected phenotypes from clinical features [142]. The difference between predicted and observed outcomes allows penetrance to be treated as a continuous, quantifiable trait in the individual. This approach enables the identification of modifiers that buffer risk, including polygenic background, environmental exposures, and developmental timing. Although not yet applied to risk gene discovery, this approach offers a clear opportunity to distinguish lowered effect size from incomplete penetrance.

Given these challenges and opportunities, the field's next phase will depend on both sample expansion and methodological diversification. Larger biobanks alone will not yield proportional gains in discovery within the intermediate frequency spectrum using standard methods, because the relationship between sample size and discovery power is intrinsically non-linear and constrained by allele frequency. While common-variant discovery continues to scale more predictably with cohort size, progress into rarer and intermediate frequencies exhibits diminishing returns unless analytical models evolve to capture these effects. Connecting rare, intermediate, and common variation into shared statistical and biological frameworks, anchored to penetrance as a quantitative trait, have the potential to advance risk gene discovery.

## Acknowledgements

This study was supported by a grant from the Seaver Foundation and the National Institute of Mental Health (NIMH; R01MH139952)

# References

1. Piñero, J. *et al.* The DisGeNET knowledge platform for disease genomics: 2019 update. *Nucleic Acids Res.* **48**, D845–D855 (2020).

2. Amberger, J. S., Bocchini, C. A., Scott, A. F. & Hamosh, A. OMIM.org: leveraging knowledge across phenotype–gene relationships. *Nucleic Acids Res.* **47**, D1038–D1043 (2019).

3. Auton, A. *et al.* A global reference for human genetic variation. *Nature* **526**, 68–74 (2015).

4. Speed, D. & Evans, D. M. Estimating disease heritability from complex pedigrees allowing for ascertainment and covariates. *Am. J. Hum. Genet.* **111**, 680–690 (2024).

5. O'Connor, L. J. & Sella, G. Principled measures and estimates of trait polygenicity. *bioRxiv* 2025.07.10.664154 (2025) doi:10.1101/2025.07.10.664154.

6. Liu, X. *et al.* Relationship between driver gene mutations and clinical pathological characteristics in older lung adenocarcinoma. *Front. Oncol.* **13**, 1275575 (2023).

7. Shah, M. *et al.* Genomic Profiling of Driver Gene Alterations in Patients With Non–Small Cell Lung Cancer, Patterns of Treatment and Impact on Survival Outcomes: A Single Center Experience of More Than 1200 Patients. *Clin. Lung Cancer* **26**, e270–e283 (2025).

8. Li, Y. R. *et al.* Rare copy number variants in over 100,000 European ancestry subjects reveal multiple disease associations. *Nat. Commun.* **11**, 255 (2020).

9. Auwerx, C. *et al.* Rare copy-number variants as modulators of common disease susceptibility. *Genome Med.* **16**, 5 (2024).

10. Goh, C. J. *et al.* Improving CNV Detection Performance in Microarray Data Using a Machine Learning-Based Approach. *Diagnostics* **14**, 84 (2023).

11. Lin, Y., Jong, Y., Huang, P. & Tsai, C. Detection of copy number variants with chromosomal microarray in 10 377 pregnancies at a single laboratory. *Acta Obstet. Gynecol. Scand.* **99**, 775–782 (2020).

12. Demidov, G. *et al.* Comprehensive reanalysis for CNVs in ES data from unsolved rare disease cases results in new diagnoses. *Npj Genomic Med.* **9**, 1–24 (2024).

13. Malwade, S., Sellgren, C. M., Vasistha, N. A. & Khodosevich, K. Rare Copy Number Variants Reveal Critical Cell Types and Periods of Brain Development in Neurodevelopmental Disorders. *Biol. Psychiatry Glob. Open Sci.* **5**, 100495 (2025).
21